\documentclass[10.5pt,compsoc]{tst}
\usepackage{graphicx}
\usepackage{footmisc}
\usepackage{subfigure}
\usepackage{url}
\usepackage{multirow}
\usepackage[noadjust]{cite}
\usepackage{amsmath,amsthm}
\usepackage{amssymb,amsfonts}
\usepackage{booktabs}
\usepackage{color}
\usepackage{ccaption}
\usepackage{booktabs}
\usepackage{float}
\usepackage{fancyhdr}
\usepackage{caption}
\usepackage{xcolor,stfloats}
\usepackage{comment}
\graphicspath{{figures/}}
\usepackage{cuted}  
\usepackage{captionhack}
\usepackage{epstopdf}
\usepackage{tabularx}
\usepackage{rotating}
\usepackage{array}
\usepackage{pifont}

\headevenname{\zihao{-5}{\textbf{\emph{Tsinghua Science and Technology, February}} 2018, 23(1): 000--000}}%
\headoddname{{\sf Xinhao Deng et al.:}\quad {\textbf{\emph{Beyond a Single Perspective: Towards a Realistic Evaluation of Website Fingerprinting Attacks}}}}%

\newtheoremstyle{mystyle}{0pt}{0pt}{\normalfont}{1em}{\bf}{}{1em}{}
\theoremstyle{mystyle}

\newcommand{\nop}[1]{}

\newcolumntype{R}[1]{>{\centering\arraybackslash}p{#1}}
\newcommand{\cmark}{\textcolor[rgb]{0.008, 0.69, 0.141}{\ding{52}}}
\newcommand{\xmark}{\textcolor[rgb]{0.71, 0.039, 0.012}{\ding{55}}}

\makeatletter
\renewcommand{\@biblabel}[1]{[#1]\hfill}
\makeatother

\begin{document}
\pagestyle{plain}




\hyphenpenalty=50000

\makeatletter
\newcommand\mysmall{\@setfontsize\mysmall{7}{9.5}}

\newenvironment{tablehere}
  {\def\@captype{table}}
  {}
\newenvironment{figurehere}
  {\def\@captype{figure}}
  {}






\begin{strip}
{\center
{\zihao{3}\textbf{
Beyond a Single Perspective: Towards a Realistic Evaluation of Website Fingerprinting Attacks
}}
\vskip 9mm}

{\center {\sf \zihao{5}
Xinhao Deng, Jingyou Chen, Linxiao Yu, Yixiang Zhang, Zhongyi Gu, Changhao Qiu, Xiyuan Zhao, Ke Xu, Qi Li$^*$
}
\vskip 5mm}
{\center \zihao{-5}{\textbf{
1.~Tsinghua University, Beijing 100084, China;  \\
2.~Southeast University, Nanjing, 211189, China;  \\
3.~National University of Defense Technology, Changsha 410073, China  \\
}}}

\vskip 5mm

\centering{
\begin{tabular}{p{160mm}}

{\zihao{-5}
\linespread{1.6667} %
\noindent
\bf{Abstract:} {\sf
Website Fingerprinting (WF) attacks exploit patterns in encrypted traffic to infer the websites visited by users, posing a serious threat to anonymous communication systems. Although recent WF techniques achieve over 90\% accuracy in controlled experimental settings, most studies remain confined to single scenarios, overlooking the complexity of real-world environments. This paper presents the first systematic and comprehensive evaluation of existing WF attacks under diverse realistic conditions, including defense mechanisms, traffic drift, multi-tab browsing, early-stage detection, open-world settings, and few-shot scenarios. Experimental results show that many WF techniques with strong performance in isolated settings degrade significantly when facing other conditions. Since real-world environments often combine multiple challenges, current WF attacks are difficult to apply directly in practice. This study highlights the limitations of WF attacks and introduces a multidimensional evaluation framework, offering critical insights for developing more robust and practical WF attacks.
}
}

\end{tabular}
}
\vskip 6mm

\vskip -3mm
\zihao{6}\end{strip}

\thispagestyle{plain}%
\thispagestyle{empty}%
\makeatother

\begin{figure}[b]
\vskip -6mm
\begin{tabular}{p{44mm}}
\toprule\\
\end{tabular}
\vskip -4.5mm
\noindent
\setlength{\tabcolsep}{1pt}
\begin{tabular}{p{1.5mm}p{79.5mm}}
$\bullet$& X. Deng, Y. Zhang, Z. Gu, X. Zhao, and Q. Li are with the Institute for Network Sciences and Cyberspace, Tsinghua University. Email: \{dengxinhao@, zhangyix24@mails., gu-zy24@mails., zhaoxy23@mails., xuke@, qli01@\}tsinghua.edu.cn.\\
$\bullet$& J. Chen and L. Yu are with the School of Cyber Science and Engineering, Southeast University. Email: \{230240004, yulinxiaobbb\}@seu.edu.cn.\\
$\bullet$& C. Qiu is with the National Key Laboratory of Information Systems Engineering, National University of Defense Technology. E-mail: qch@nudt.edu.cn.\\
$\bullet$& K. Xu is with the Department of Computer Science, Tsinghua University. Email: xuke@tsinghua.edu.cn.\\

\end{tabular}
\end{figure}\zihao{5}

\section{Introduction}
\label{s:introduction}
\noindent
Website Fingerprinting (WF) attacks have emerged as an important and active research area in anonymous communication over the Internet~\cite{tor2024users, dingledine2004tor}. 
By analyzing encrypted traffic features such as packet size, direction, and timing patterns, adversaries can infer the websites visited by users, thereby undermining the privacy guarantees of encrypted communication. 
Over the past decade, WF techniques have evolved from statistical learning approaches to deep learning models, demonstrating steadily improving performance under controlled experimental conditions~\cite{wang2014effective, hayes2016k, wang2018deep, xu2018multi, rimmer2018automated}. 
Existing research typically frames WF attacks as a classification problem, employing machine learning methods (e.g., support vector machines, random forests) or deep learning architectures (e.g., convolutional neural networks, Transformers) to automatically extract traffic features. However, most reported results achieve high accuracy only within isolated experimental settings, lacking systematic evaluation under realistic conditions.

In practice, multiple challenges significantly affect the performance of WF attacks. Juarez et al.~\cite{juarez2014critical} have shown that attack accuracy decreases substantially when tested against larger website sets~\cite{wang2020high}, diverse Tor versions~\cite{torversion}, more complex user behaviors~\cite{deng2023robust, jin2023transformer, guan2021bapm} and temporal drift~\cite{rimmer2018automated, sirinam2019triplet}. In other words, many high-accuracy results rely on unrealistic assumptions, such as single-tab browsing, static website content, or consistency between training and testing data~\cite{juarez2014critical}. Moreover, adversaries are often assumed to possess complete knowledge of a user’s browsing scope, an assumption that rarely holds in open network environments~\cite{jansen2024measurement}. Studies based on real Tor traffic further reveal that even state-of-the-art models experience dramatic performance degradation in more realistic open-world scenarios. For example, when the monitored set includes only five popular websites, accuracy can exceed 95\%; yet when expanded to 25 websites, accuracy drops below 80\%~\cite{cherubin2022online}. This illustrates the practical difficulty of reliably identifying and monitoring large-scale website visits.

\begin{figure}[t]
    \centering
    \includegraphics[width=\linewidth]{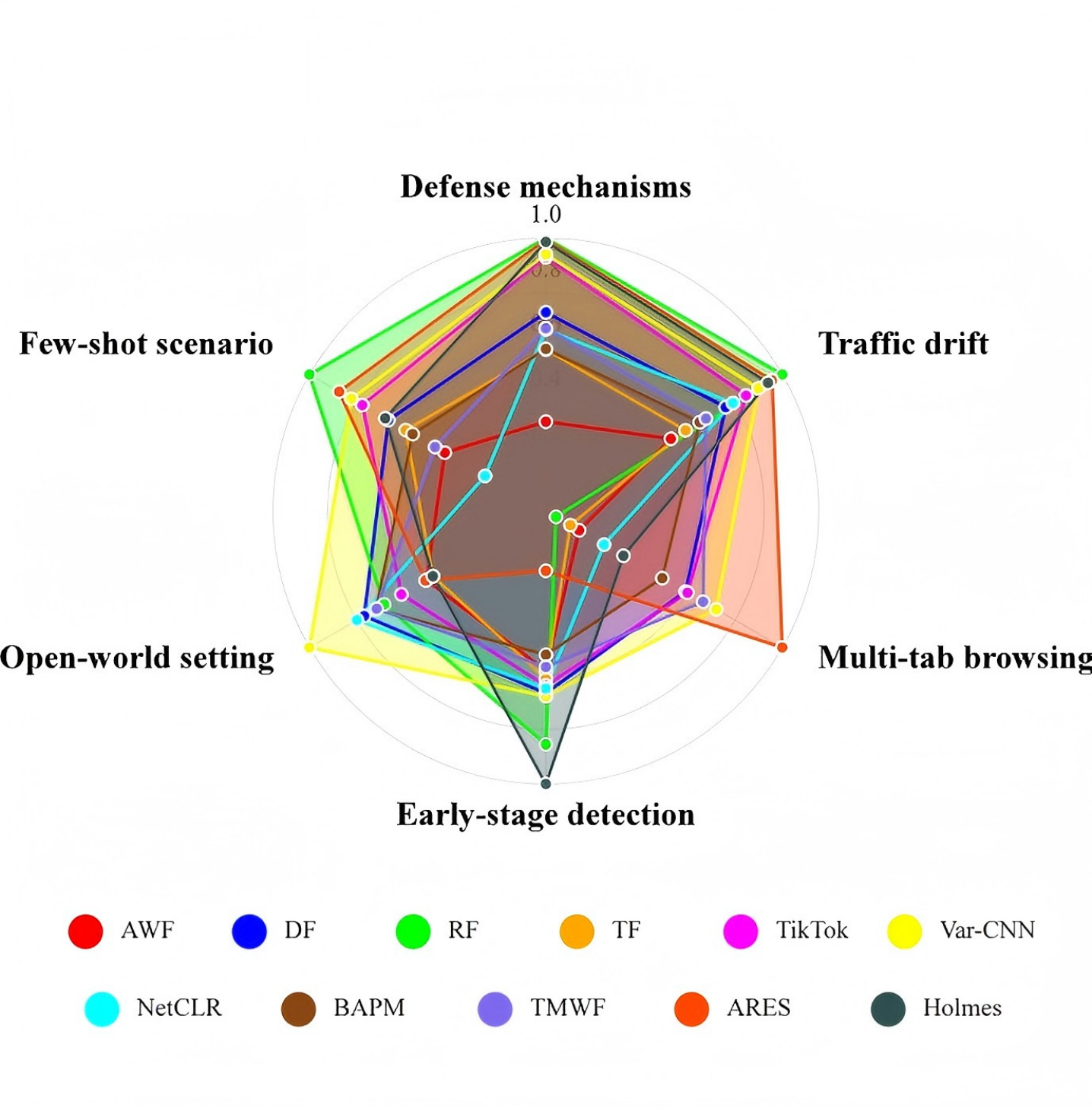}
    \caption{Comparative evaluation of website fingerprinting attacks across key real-world scenarios.}
    \label{fig:comp-wfp}
\end{figure}

To address these limitations, this paper proposes a more comprehensive and realistic WF attack evaluation framework. We identify and categorize six key challenges: (i) defense mechanisms, (ii) traffic drift, (iii) multi-tab browsing, (iv) early-stage detection, (v) open-world scenarios, and (vi) the few-shot settings. Based on these challenges, we construct a unified experimental evaluation system that integrates them into WF research and systematically analyzes mainstream methods under complex conditions. Experimental results in Figure~\ref{fig:comp-wfp} demonstrate that many WF techniques achieving strong performance in single scenarios suffer substantial accuracy degradation in other contexts. This finding underscores the limitations of current methods and highlights the need for future research to conduct comprehensive evaluations in diverse and dynamic environments, thereby advancing the development of more practical and robust WF attack techniques.

\section{Background \& Related Works}
\label{s:background}

\subsection{WF Attacks}

\noindent \textbf{Traditional Approaches}. 
Early website fingerprinting (WF) attacks primarily relied on handcrafted traffic features in combination with traditional machine learning classifiers. In these approaches, attackers first collected encrypted traffic traces from various websites and then extracted statistical features such as sequence length, packet size distribution, traffic directionality, and inter-packet timing. Liberatore and Levine~\cite{liberatore2006inferring} pioneered the use of packet-size frequency histograms, while Panchenko et al.~\cite{panchenko2011website} improved accuracy through refined distance statistics and support vector machines (SVMs). Later, Cai et al.~\cite{cai2012editdistance} introduced edit-distance and template-matching techniques.  
These methods generally involved manual feature engineering, followed by classification using algorithms such as $k$-nearest neighbors ($k$-NN), Naïve Bayes, SVMs, or random forests. Although they achieved relatively high accuracy (often exceeding 80–90\%) in small-scale closed-world settings, their performance was highly sensitive to feature selection and parameter tuning. Moreover, their effectiveness plateaued in large-scale or more complex environments, limiting their practical deployment.

\noindent \textbf{Deep Learning Approaches}. 
With the rise of deep learning, researchers increasingly shifted toward automated feature extraction, thereby reducing reliance on manual engineering. A variety of neural architectures have been proposed for WF attacks. Sirinam et al.~\cite{sirinam2018deep} introduced Deep Fingerprinting, which leverages convolutional neural networks (CNNs) to directly learn from raw packet sequences. Building on this, Rimmer et al.~\cite{rimmer2018automated} and others explored residual and recurrent neural architectures. These models automatically capture discriminative traffic patterns and have achieved remarkable accuracy in closed-world settings, often exceeding 95\%. For example, Sirinam’s CNN model reported over 98\% accuracy on undefended Tor traffic. Subsequent variants, such as Var-CNN~\cite{bhat2019varcnn} and Triplet Fingerprinting~\cite{sirinam2019triplet}, further enhanced temporal robustness and inter-class separability.  
Despite these advances, deep learning models generally require large-scale training datasets and relatively stable operating conditions. Their performance tends to degrade under data scarcity or distribution shifts—issues that will be revisited later in this study.

These specialized efforts represent significant progress toward practical WF attacks but often optimize for isolated challenges. Their limited adaptability under complex, real-world conditions motivates the broader evaluation undertaken in this work.

\subsection{Focal Aspects of Attacks}
We categorize WF attacks according to their primary focus and the challenges they address, as summarized in Table~\ref{tab:attack_focus}.

\begin{table*}[htbp]
\centering
\renewcommand{\arraystretch}{1.3} 
\setlength{\tabcolsep}{5pt} 

\begin{tabularx}{\textwidth}{l c | X | *{6}{R{.6cm}}}
\toprule
\textbf{Attack} & 
\textbf{Year} & 
\textbf{Description} & 
\rotatebox{60}{\textbf{Defense}} & 
\rotatebox{60}{\textbf{Traffic Drift}} & 
\rotatebox{60}{\textbf{Multi-tab}} & 
\rotatebox{60}{\textbf{Early-Stage}} & 
\rotatebox{60}{\textbf{Open-World}} &
\rotatebox{60}{\textbf{Few-Shot}}  \\
\midrule
AWF~\cite{rimmer2018automated} & 2018 & Deep learning–based WF with automated hyperparameter tuning for optimal model selection. 
& \xmark & \cmark & \xmark & \xmark & \cmark & \xmark \\
DF~\cite{sirinam2018deep} & 2018 & CNN-based approach to capture richer traffic features and mitigate overfitting. 
& \cmark & \xmark & \xmark & \xmark & \cmark & \xmark \\
Var-CNN~\cite{bhat2019varcnn} & 2019 & Ensemble of softmax outputs from packet timing and direction, reducing training data requirements. 
& \cmark & \cmark & \xmark & \xmark & \cmark & \cmark \\
Tik-Tok~\cite{rahman2019tik} & 2019 & Histogram-based burst features to improve resilience against padding defenses. 
& \cmark & \cmark & \xmark & \xmark & \cmark & \xmark \\
TF~\cite{sirinam2019triplet} & 2019 & Triplet-network feature extractor for robust and transferable traffic representations. 
& \cmark & \cmark & \xmark & \xmark & \cmark & \cmark \\
BAPM~\cite{guan2021bapm} & 2021 & Attention mechanism to segment multi-tab traffic into single-tab flows for classification. 
& \xmark & \xmark & \cmark & \xmark & \cmark & \xmark \\
ARES~\cite{deng2023robust} & 2023 & Multi-label framework for tab-level traffic division and website recognition without prior knowledge. 
& \cmark & \cmark & \cmark & \xmark & \cmark & \xmark \\
NetCLR~\cite{bahramali2023realistic} & 2023 & Burst-based augmentation to simulate diverse network conditions and enhance robustness to drift. 
& \cmark & \cmark & \xmark & \xmark & \cmark & \cmark \\
TMWF~\cite{jin2023transformer} & 2023 & Transformer-based WF for multi-tab traffic with calibrated metrics for realistic evaluation. 
& \xmark & \xmark & \cmark & \xmark & \cmark & \xmark \\
RF~\cite{shen2023rf} & 2023 & Temporal-Aware Module (TAM) jointly encodes time and direction to improve defense robustness. 
& \xmark & \cmark & \xmark & \xmark & \cmark & \xmark \\
Holmes~\cite{deng2024holmes} & 2024 & Early-stage WF using supervised contrastive learning to extract website features from limited traffic. 
& \cmark & \cmark & \cmark & \cmark & \cmark & \xmark \\
\bottomrule
\end{tabularx}

\caption{Summary of representative WF attacks and their addressed challenges.}
\label{tab:attack_focus}
\end{table*}

\textbf{AWF}~\cite{rimmer2018automated}. The Automated Website Fingerprinting (AWF) attack provides the first systematic exploration of deep learning algorithms, such as CNNs and LSTMs, to automatically extract website traffic features and select hyperparameters. It evaluates models on a large open-world dataset to demonstrate competitive performance. Moreover, it re-evaluates traffic from different time periods to assess robustness against concept drift.

\textbf{DF}~\cite{sirinam2018deep}. DF employs a deep network consisting of multiple consecutive CNN blocks, in contrast to AWF’s single CNN block. This design enables more effective feature extraction. DF further incorporates modern deep learning techniques, such as \textit{batch normalization} and \textit{hyperparameter variation}, to mitigate overfitting and capture richer traffic information. Consequently, DF achieves stronger performance against defenses such as WTF-PAD.

\textbf{Var-CNN}~\cite{bhat2019varcnn}. Var-CNN observes that packet sequences exhibit long-term dependencies, which render RNNs and LSTMs prohibitively expensive to train when modeling global information. To address this, Var-CNN adopts \textit{dilated convolutions}, which capture temporal relationships in packet sequences without increasing training overhead. By ensembling the softmax outputs of packet timing and direction, Var-CNN achieves competitive performance with less training data than previous approaches.

\textbf{Tik-Tok}~\cite{rahman2019tik}. Rather than packet-level features, Tik-Tok extracts burst-level timing features, such as the timestamp of the \textit{median packet in a burst} (MED). It then constructs histograms of these features, multiplied by burst direction. This directional histogram captures both local and global burst characteristics and is resilient to padding-based defenses, since packet insertion minimally affects burst-level features like MED.

\textbf{TF}~\cite{sirinam2019triplet}. Sirinam et al. applied a \textit{Triplet Network}~\cite{schroff2015facenet}, which minimizes intra-class distances while maximizing inter-class distances, to learn discriminative traffic representations. During classification, the triplet network first encodes traces into embedding vectors, which are then passed to a pre-trained classifier. This decoupling of feature extraction and classification reduces retraining costs when incorporating new websites or traffic collected after significant time intervals.

\textbf{BAPM}~\cite{guan2021bapm}. BAPM focuses on multi-tab WF attacks. Given a trace generated by sequentially requesting multiple tabs, BAPM divides the trace into blocks, each representing partial traffic features. It then employs an \textit{attention mechanism}~\cite{vaswani2017attention} to model inter-block relationships. Blocks with higher attention scores are grouped to identify the originating website. Although effective at separating mixed traffic, BAPM requires prior knowledge of the number of tabs, which limits its practicality.

\textbf{ARES}~\cite{deng2023robust}. ARES reframes multi-tab WF as a multi-label classification problem. Instead of a single model, it deploys a dedicated model, Trans-WF, for each website. ARES aggregates the predictions of all Trans-WFs to determine which websites are visited. It employs a \textit{multi-head Top-$m$ attention mechanism} to strengthen local patterns of the same website and weaken those of different websites. This approach enables robust detection across varying numbers of tabs, without requiring prior knowledge of tab counts. Adding a new website only requires training a new Trans-WF, improving scalability and practical usability.

\textbf{NetCLR}~\cite{bahramali2023realistic}. NetCLR introduces data augmentation to simulate traffic under diverse network conditions. For each burst, it generates variants reflecting dynamic content and fluctuating Tor bandwidth by modifying, inserting, or merging bursts. During training, NetCLR uses pre-training on unlabeled traces to learn robust representations, followed by fine-tuning with labeled traces. Experiments show that training on more challenging (interior) traces improves robustness to network variability and enhances generalization.

\textbf{TMWF}~\cite{jin2023transformer}. TMWF integrates the Transformer~\cite{vaswani2017attention} as its classification module while adopting DF as the feature extractor. For websites outside the monitored set, it assigns a \textit{no-tab} label to filter irrelevant predictions. Additionally, TMWF proposes new metrics to recalibrate accuracy, precision, and recall when unmonitored websites are present, reflecting realistic attacker goals. The method relies on a predefined parameter $N$ (maximum number of tabs), but setting $N$ sufficiently large improves practicality.

\textbf{RF}~\cite{shen2023rf}. RF measures information leakage across different traffic features with and without defenses. Based on this analysis, it proposes a novel representation called TAM, which encodes packet counts and directions within fixed time slots. TAM tolerates obfuscation such as padding and delays, making it robust to multiple defenses. The model integrates 2-D and 1-D CNNs with $2\times2$ max-pooling and replaces fully connected layers with GAP~\cite{lin2013network} layers to reduce overfitting. RF achieves superior open-world performance, particularly under defensive settings.

\textbf{Holmes}~\cite{deng2024holmes}. In real-world settings, only partial traces may be observable. Models trained on complete traces often degrade severely under such conditions. Holmes targets this challenge by identifying websites using \textit{early-stage traffic} collected during the initial page load. The authors show that early-stage traffic contains sufficient information for classification. Holmes employs spatial and temporal data augmentation to enrich early-stage traces and applies \textit{supervised contrastive learning}~\cite{khosla2020supervised} to align early and complete traces in latent space. During deployment, it adaptively classifies traces once confidence surpasses a threshold, thereby improving robustness against various defenses.

\section{Threat Model \& Problem Statement}

\subsection{Threat Model}

\begin{figure}
    \centering
    \includegraphics[width=\linewidth]{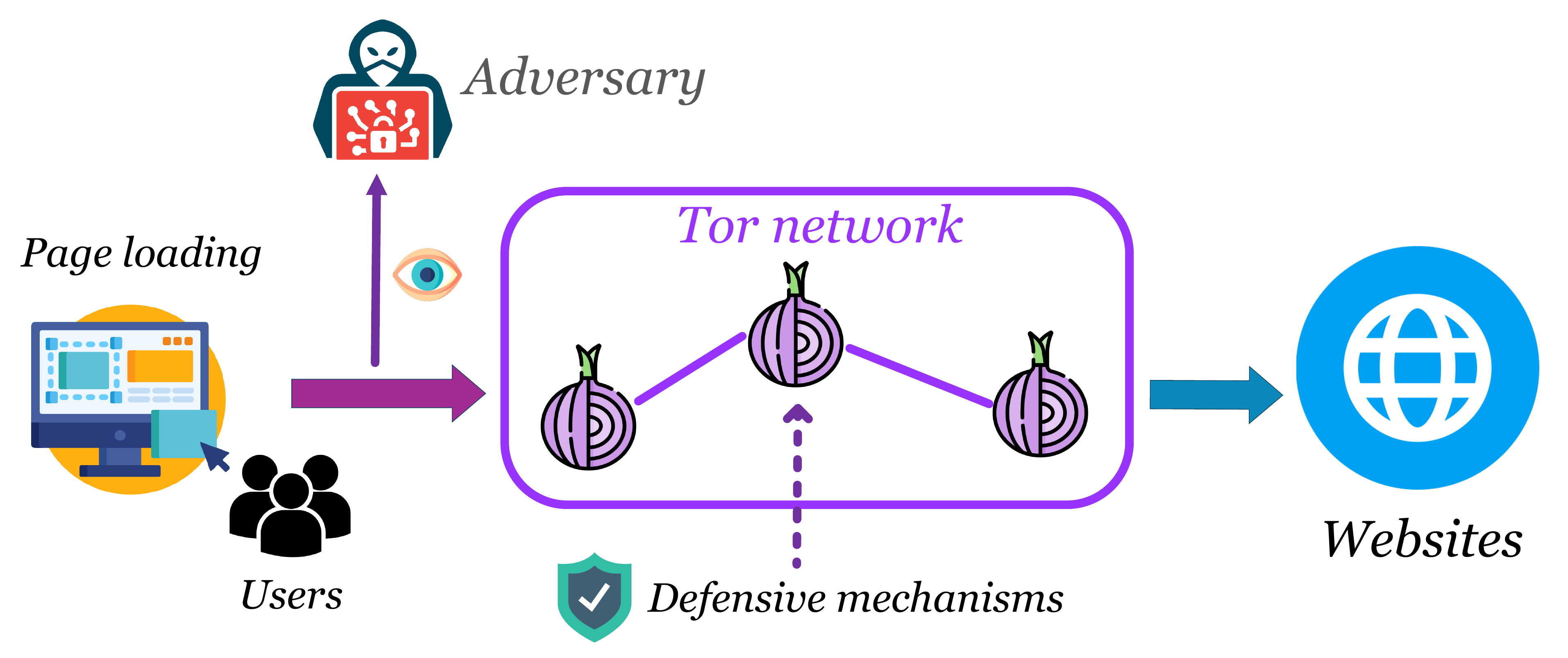}
    \caption{Website Fingerprinting Attack Model}
    \label{fig:Attack_Model}
\end{figure}

Figure~\ref{fig:Attack_Model} presents the threat model for website fingerprinting (WF) attacks, which aim to infer visited websites by analyzing encrypted traffic patterns, thereby undermining anonymous communication. We assume a \emph{local passive adversary}—such as an Internet Service Provider (ISP) or a compromised Tor entry (Guard) relay—who can monitor all traffic between the user and the Tor network~\cite{juarez2014critical}. The adversary cannot access plaintext content but can collect metadata (packet timing, direction, and size), and seeks to reconstruct the user's browsing activity from these observable traces.

In a closed-world setting, WF is formulated as a multi-class classification problem over a predefined set of monitored websites: the adversary collects traffic samples per site and trains a classifier mapping observed traces to one of these classes. In an open-world setting, an additional ``other'' class represents unmonitored sites, transforming the task into an $(N+1)$-class classification or, equivalently, a detection problem in which the adversary must both identify monitored sites and reject unknowns. In both cases, the adversary's objective is to maximize identification accuracy under given constraints.

\subsection{Problem Statement}

Although WF attacks succeed when websites yield statistically distinct traffic and sufficient training data are available, several practical factors significantly reduce performance:

\begin{itemize}
\item \textbf{Defense mechanisms.}
Traffic obfuscation techniques—such as adaptive padding, random delays, and other shaping defenses—diminish discriminative features and introduce noise, often causing sharp accuracy drops unless countered with tailored methods~\cite{wang2017walkie}.
\item \textbf{Traffic drift.}
Temporal variations in site content and network conditions induce distribution shifts; models trained on outdated data degrade unless regularly updated or adapted~\cite{yang2021cade}.
\item \textbf{Multi-tab browsing.}
Concurrent page loads interleave flows, converting single-trace classification into a de-noising or multi-label recognition task~\cite{xu2018multi}.
\item \textbf{Early-stage detection.}
Partial captures (e.g., initial load segments) contain limited and unstable signals, heightening uncertainty and reducing recognition accuracy~\cite{deng2024holmes}.
\item \textbf{Open-world uncertainty.}
The base-rate problem arises when most traffic belongs to unmonitored sites; even low false-positive rates may yield many errors, requiring careful thresholding and calibration~\cite{wang2020high}.
\item \textbf{Few-shot setting.}
Limited labeled samples for new or drifting sites challenge both deep and statistical models; transfer learning, alignment, and augmentation provide partial remedies, but robustness under compounded constraints remains uncertain~\cite{oh2021gandalf}.
\end{itemize}
\section{Attack Analysis}

In this section, we evaluate the performance of existing WF attacks across six practical scenarios: defense mechanisms, traffic drift, multi-tab browsing, early-stage detection, open-world and few-shot setting.

\subsection{Dataset Construction}

We construct six evaluation datasets to assess WF methods comprehensively. The defense dataset covers WTF-PAD, Front, and Walkie-Talkie defenses with traffic from 95 websites. To examine temporal robustness, a longitudinal dataset of 102 websites spanning March 2024 to July 2025 yields 145,896 traces, with models trained on day-0 traffic and tested across later checkpoints. Multi-tab browsing is simulated with 2–5 tab sessions over 100 websites, producing 58,000 traces for evaluation. For early-stage detection, traffic from 95 Alexa sites is segmented at 20\%, 40\%, 60\%, and 80\% load intervals. Open-world evaluation employs GTT23, with 100 monitored sites and large-scale unmonitored traces disjoint between training and testing. Finally, few-shot scenarios fine-tune day-0 models with 10 traces per class from drifted checkpoints (day 270 and 480), testing adaptability under limited supervision.

\subsection{Evaluation under Defense Mechanisms}

\begin{table}[t]
  \centering
  \setlength{\belowcaptionskip}{0.1cm}
  \caption{Comparison of different methods under Defense Mechanisms.}
  \label{tab:defense_evaluation}

  \begin{tabular}{lcccc}
    \toprule
    {Method} & {Acc.} & {Prc.} & {Rec.} & {F1} \\
    \midrule
    AWF     & 0.3198 & 0.3203 & 0.3199 & 0.3141 \\
    DF      & 0.7042 & 0.7053 & 0.7066 & 0.6982 \\
    Var-CNN & 0.9007 & 0.9130 & 0.8995 & 0.9019 \\
    TikTok  & 0.8910 & 0.8971 & 0.8918 & 0.8922 \\
    TF      & 0.5767 & 0.5772 & 0.5693 & 0.5673 \\
    BAPM    & 0.5750 & 0.6023 & 0.5718 & 0.5697 \\
    ARES    & 0.9530 & 0.9560 & 0.9537 & 0.9542 \\
    NetCLR  & 0.6490 & 0.6424 & 0.6403 & 0.6382 \\
    TMWF    & 0.6550 & 0.6580 & 0.6486 & 0.6423 \\
    RF      & \textbf{0.9582} & \textbf{0.9605} & \textbf{0.9591} & \textbf{0.9593} \\
    Holmes  & 0.9458 & 0.9469 & 0.9458 & 0.9458 \\
    \bottomrule
  \end{tabular}%
\end{table}

For the evaluation of adversarial defenses, we construct a defense dataset that integrates three representative mechanisms: WTF-PAD, Front, and Walkie-Talkie. The dataset comprises 95 websites, each contributing approximately 1,000 traces. For each trace, one of the three defenses is applied with equal probability. This mixed-defense dataset enables the evaluation of robustness against composite defense strategies.

As shown in Table~\ref{tab:defense_evaluation}, classifier accuracy varies notably depending on the feature representation. When using only packet direction sequences, accuracy ranges from 31.98\% to 70.42\%. Incorporating both packet direction and timestamp information substantially improves performance, yielding accuracies of 89.10\%–90.07\%. Further enhancement is achieved through feature aggregation, with results reaching 94.58\%–95.82\%.

This performance trend can be explained by the characteristics of the defense mechanisms. Specifically, they introduce a large number of chaff packets into the traces, which severely degrades the discriminative power of packet direction features but has only a limited effect on timestamp information. Moreover, aggregating features through statistical analysis of traffic within fixed-length time windows further mitigates temporal distortions, thereby improving robustness.
\subsection{Evaluation under Traffic Drift}

\begin{table}[t]
\centering
\setlength{\belowcaptionskip}{0.1cm}
\caption{Performance of WF methods under traffic drift.}
\label{tab:traffic_drift}
\begin{tabular}{lcccc}
\toprule
\textbf{Method} & \textbf{Acc.} & \textbf{Prc.} & \textbf{Rec.} & \textbf{F1-Score} \\
\midrule
AWF & 0.3706 & 0.3715 & 0.3704 & 0.3602 \\
DF & 0.5247 & 0.5302 & 0.5244 & 0.5179 \\
Var-CNN & 0.6188 & 0.6331 & 0.6185 & 0.6122 \\
TikTok & 0.5851 & 0.5971 & 0.5848 & 0.5767 \\
TF & 0.4101 & 0.4139 & 0.4098 & 0.4025 \\
BAPM & 0.4519 & 0.4755 & 0.4517 & 0.4431 \\
ARES & 0.6595 & 0.6750 & 0.6591 & 0.6517 \\
NetCLR & 0.5466 & 0.5491 & 0.5463 & 0.5387 \\
TMWF & 0.4676 & 0.4908 & 0.4674 & 0.4612 \\
RF & \textbf{0.6911} & \textbf{0.7073} & \textbf{0.6906} & \textbf{0.6810} \\
Holmes & 0.6458 & 0.6594 & 0.6454 & 0.6397 \\
\bottomrule
\end{tabular}
\end{table}

To evaluate robustness under temporal distribution shifts, all models are trained and validated on traffic collected from day-0, and subsequently tested on uniformly sampled traces from later checkpoints. 

The results are presented in Table~\ref{tab:traffic_drift}. 
Overall, traffic drift leads to a noticeable degradation across all methods, underscoring the inherent difficulty of sustaining stable classification performance as websites evolve. 
All approaches rely on directional sequences to varying extents. Among them, RF, ARES, and Holmes, which incorporate more comprehensive feature representations, demonstrate the strongest robustness, with RF achieving the highest accuracy of 0.6911. 
TikTok and Var-CNN, which enhance directional sequences with temporal information, achieve moderate gains but remain inferior to RF and ARES. 
In contrast, AWF and TF, which rely solely on directional sequences without incorporating richer features, perform the weakest and exhibit the largest accuracy drops. 
These findings suggest that while directional sequences provide a common foundation, the robustness of WF models under traffic drift is highly dependent on the richness and diversity of the employed feature representations.

\subsection{Evaluation under Multi-Tab Browsing}

To evaluate robustness under realistic browsing behaviors, we construct four multi-tab datasets (2-tab, 3-tab, 4-tab, and 5-tab) by visiting 100 monitored websites over Tor in a closed-world setting. Each website contributes equally, and all datasets are of identical size. To simulate concurrent browsing, 25\% of the traces in each dataset are randomly merged into a mixed dataset, resulting in 46,980 training traces, 5,220 validation traces, and 5,800 test traces.

As shown in Table~\ref{tab:multitab}, most methods achieve only moderate performance, with AP@avg ranging from 0.03 to 0.57, P@avg generally below 0.40, and AUC@avg between 0.50 and 0.81. The multi-tab setting introduces significant noise and overlapping flows, which obscure website-specific patterns. In contrast, ARES effectively mitigates this challenge by leveraging explicit multi-label classification, multi-level traffic aggregation, and top-m attention. Overall, multi-tab browsing—essentially a multi-label task—remains a major obstacle for practical website fingerprinting.

\begin{table}[t]
\centering
\setlength{\belowcaptionskip}{0.1cm}
\caption{Performance under Multi-Tab Browsing}
\label{tab:multitab}
\begin{tabular}{lccc}
\toprule
\textbf{Method} & \textbf{AUC@avg} & \textbf{P@avg} & \textbf{AP@avg} \\
\midrule
AWF     & 0.6049 & 0.0950 & 0.1104 \\
DF      & 0.7610 & 0.3257 & 0.4630 \\
Var-CNN & 0.8100 & 0.4110 & 0.5650 \\
TikTok  & 0.7689 & 0.3295 & 0.4693 \\
TF      & 0.5433 & 0.0701 & 0.0814 \\
BAPM    & 0.7790 & 0.2869 & 0.3858 \\
ARES    & \textbf{0.9432} & \textbf{0.6439} & \textbf{0.7839} \\
NetCLR  & 0.6594 & 0.1535 & 0.1930 \\
TMWF    & 0.8417 & 0.3908 & 0.5211 \\
RF      & 0.4972 & 0.0339 & 0.0336 \\
Holmes  & 0.6405 & 0.1939 & 0.2571 \\
\bottomrule
\end{tabular}
\end{table}

\subsection{Evaluation with Early-Stage Detection}

\begin{table*}[!htb]
\centering
\setlength{\belowcaptionskip}{0.1cm}
\caption{Comparison of Different Methods at Different Early Stages}
\label{tab:Early_Stage_evaluation}
\resizebox{\textwidth}{!}{
\begin{tabular}{lcccccccccccccccc}
\toprule
\multirow{2}{*}{\textbf{Method}} 
& \multicolumn{4}{|c|}{\textbf{20\% loaded}} 
& \multicolumn{4}{c|}{\textbf{40\% loaded}} 
& \multicolumn{4}{c|}{\textbf{60\% loaded}}
& \multicolumn{4}{c}{\textbf{80\% loaded}} \\ 
\cline{2-5} \cline{6-9} \cline{10-13} \cline{14-17}
& \multicolumn{1}{|c}{\textbf{Acc.}} 
& \multicolumn{1}{c}{\textbf{Prc.}} 
& \multicolumn{1}{c}{\textbf{Rec.}}
& \multicolumn{1}{c|}{\textbf{F1}}
& \multicolumn{1}{c}{\textbf{Acc.}} 
& \multicolumn{1}{c}{\textbf{Prc.}} 
& \multicolumn{1}{c}{\textbf{Rec.}}
& \multicolumn{1}{c|}{\textbf{F1}}
& \multicolumn{1}{c}{\textbf{Acc.}} 
& \multicolumn{1}{c}{\textbf{Prc.}} 
& \multicolumn{1}{c}{\textbf{Rec.}}
& \multicolumn{1}{c|}{\textbf{F1}}
& \multicolumn{1}{c}{\textbf{Acc.}} 
& \multicolumn{1}{c}{\textbf{Prc.}} 
& \multicolumn{1}{c}{\textbf{Rec.}}
& \multicolumn{1}{c}{\textbf{F1}} \\
\midrule
AWF 
& \multicolumn{1}{|c}{0.0950} & 0.3072 & 0.0953 & 0.1185 
& \multicolumn{1}{|c}{0.3200} & 0.4678 & 0.3201 & 0.3465 
& \multicolumn{1}{|c}{0.6335} & 0.6834 & 0.6334 & 0.6411 
& \multicolumn{1}{|c}{0.8717} & 0.8803 & 0.8716 & 0.8691 \\
DF 
& \multicolumn{1}{|c}{0.1289} & 0.3780 & 0.1298 & 0.1481
& \multicolumn{1}{|c}{0.4070} & 0.5670 & 0.4073 & 0.4293
& \multicolumn{1}{|c}{0.7457} & 0.7835 & 0.7457 & 0.7475
& \multicolumn{1}{|c}{0.9294} & 0.9392 & 0.9295 & 0.9282\\
Var-CNN 
& \multicolumn{1}{|c}{0.1225} & 0.4267 & 0.1227 & 0.1431
& \multicolumn{1}{|c}{0.4323} & 0.6590 & 0.4322 & 0.4624
& \multicolumn{1}{|c}{0.7655} & 0.8102 & 0.7655 & 0.7658
& \multicolumn{1}{|c}{0.9371} & 0.9434 & 0.9372 & 0.9354\\
TikTok 
& \multicolumn{1}{|c}{0.1172} & 0.3751 & 0.1174 & 0.1325
& \multicolumn{1}{|c}{0.3721} & 0.5288 & 0.3719 & 0.3848
& \multicolumn{1}{|c}{0.7044} & 0.7498 & 0.7043 & 0.7080
& \multicolumn{1}{|c}{0.9188} & 0.9262 & 0.9188 & 0.9163\\
TF 
& \multicolumn{1}{|c}{0.0933} & 0.3328 & 0.0938 & 0.1083
& \multicolumn{1}{|c}{0.3291} & 0.5099 & 0.3295 & 0.3583
& \multicolumn{1}{|c}{0.6877} & 0.7369 & 0.6876 & 0.6947
& \multicolumn{1}{|c}{0.9095} & 0.9228 & 0.9095 & 0.9081\\
BAPM 
& \multicolumn{1}{|c}{0.0645} & 0.2613 & 0.0647 & 0.0805
& \multicolumn{1}{|c}{0.2090} & 0.4022 & 0.2092 & 0.2334
& \multicolumn{1}{|c}{0.5556} & 0.6449 & 0.5553 & 0.5694
& \multicolumn{1}{|c}{0.8613} & 0.8862 & 0.8613 & 0.8629\\
ARES 
& \multicolumn{1}{|c}{0.0265} & 0.0633 & 0.0273 & 0.0263
& \multicolumn{1}{|c}{0.0814} & 0.1382 & 0.0751 & 0.0742
& \multicolumn{1}{|c}{0.1894} & 0.2902 & 0.1743 & 0.1834
& \multicolumn{1}{|c}{0.4867} & 0.5334 & 0.4708 & 0.4552\\
NetCLR 
& \multicolumn{1}{|c}{0.1242} & 0.3314 & 0.1246 & 0.1400
& \multicolumn{1}{|c}{0.4093} & 0.5325 & 0.4095 & 0.4190
& \multicolumn{1}{|c}{0.7167} & 0.7546 & 0.7167 & 0.7160
& \multicolumn{1}{|c}{0.9212} & 0.9298 & 0.9212 & 0.9195\\
TMWF 
& \multicolumn{1}{|c}{0.0911} & 0.2908 & 0.0918 & 0.1029
& \multicolumn{1}{|c}{0.2822} & 0.4401 & 0.2826 & 0.2977
& \multicolumn{1}{|c}{0.6353} & 0.6900 & 0.6350 & 0.6348
& \multicolumn{1}{|c}{0.8963} & 0.9127 & 0.8962 & 0.8937\\
RF 
& \multicolumn{1}{|c}{0.2474} & 0.5748 & 0.2477 & 0.2793
& \multicolumn{1}{|c}{0.7029} & 0.8068 & 0.7025 & 0.7163
& \multicolumn{1}{|c}{0.9175} & 0.9272 & 0.9175 & 0.9170
& \multicolumn{1}{|c}{0.9742} & 0.9769 & 0.9742 & 0.9740\\
Holmes 
& \multicolumn{1}{|c}{\textbf{0.5132}} & \textbf{0.6673} & \textbf{0.5129} & \textbf{0.5292}
& \multicolumn{1}{|c}{\textbf{0.9048}} & \textbf{0.9090} & \textbf{0.9046} & \textbf{0.9020}
& \multicolumn{1}{|c}{\textbf{0.9661}} & \textbf{0.9676} & \textbf{0.9661} & \textbf{0.9659}
& \multicolumn{1}{|c}{\textbf{0.9784}} & \textbf{0.9787} & \textbf{0.9784} & \textbf{0.9784}\\
\bottomrule
\end{tabular}
}
\end{table*}

For the early-stage evaluation, we constructed a benchmark dataset comprising traces collected from 95 monitored websites listed in Alexa’s top rankings, with each site contributing more than 1,000 traces. Packet timestamps were used to segment each trace into four progressive loading intervals: 20\%, 40\%, 60\%, and 80\% of the total load time. All major attack models were evaluated across these intervals.

As summarized in Table~\ref{tab:Early_Stage_evaluation}, Holmes demonstrates a clear advantage even at the very early stage of 20\% loading, achieving an F1-score of 52.92\%, substantially higher than RF (27.93\%) and ARES (2.63\%). At the 40\% loading interval, Holmes nearly attains its full-stage performance, with an F1-score of 90.20\% and accuracy, precision, and recall all approaching those observed at complete loading. At later stages (60\%–80\%), Holmes continues to lead, achieving F1-scores of 96.59\% and 97.84\%, respectively.

These results highlight that Holmes not only achieves high recognition performance but also maintains a practical advantage by enabling earlier and more reliable detection compared with competing approaches.
\subsection{Open-World Evaluation}

\begin{figure}
    \centering
    \includegraphics[width=1\linewidth]{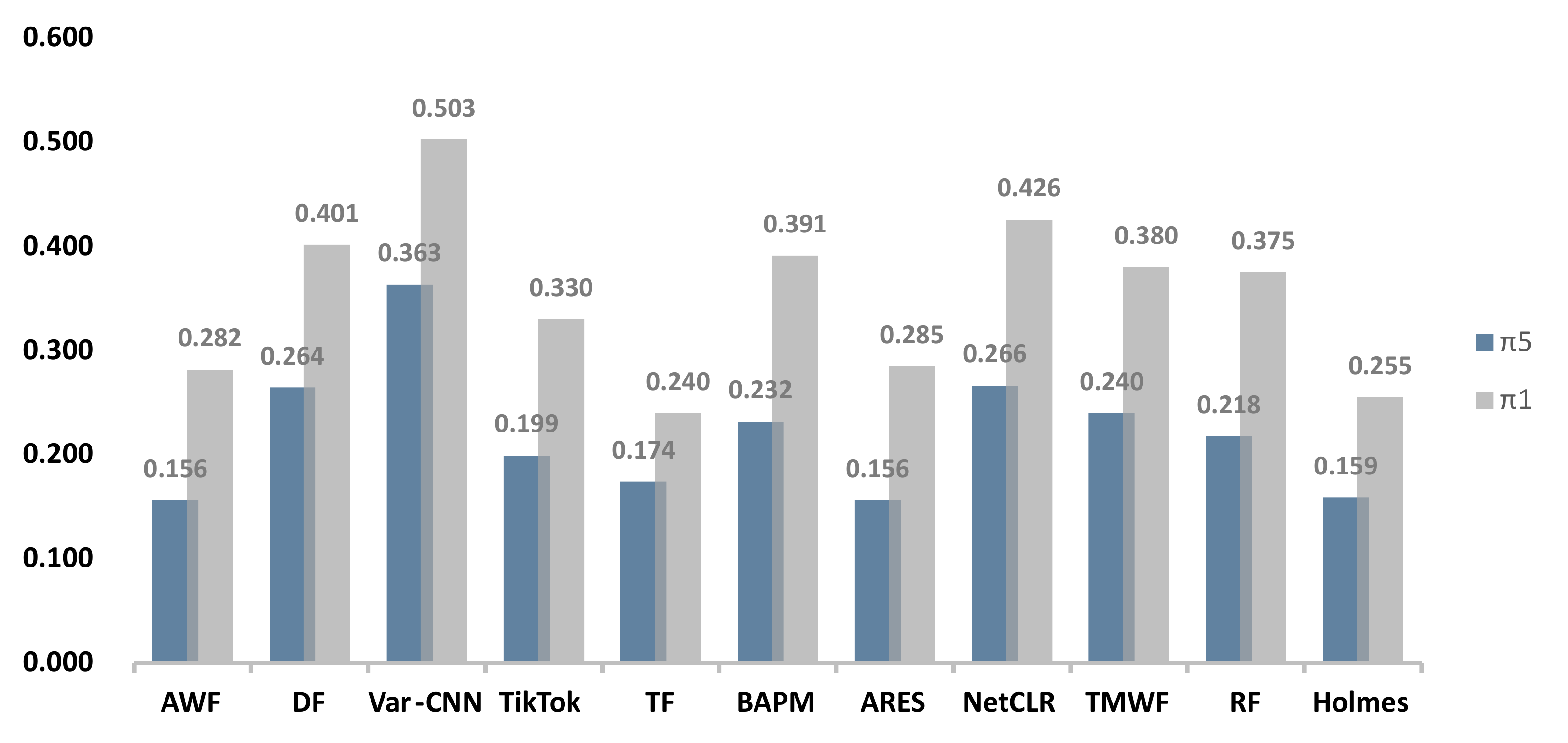} 
    \caption{Comparison of different methods in the open world}
    \label{fig:Attack Model}
\end{figure}

For open-world evaluation, we adopt the GTT23 dataset, which contains 100 monitored websites with approximately 100 traces each, along with 5,000 unmonitored traces for training. The test set includes 100 traces per monitored website and 20,000 unmonitored traces, with no overlap between the training and test websites. This design enables evaluation under realistic traffic conditions.

As shown in Figure~\ref{fig:Attack Model}, the $\pi_1$ values range from 24.0\% to 50.3\%, while $\pi_5$ is notably lower, with the best-performing method reaching only 36.3\%. Among all approaches, single-tab attacks such as DF, Var-CNN, and NetCLR consistently outperform multi-tab counterparts. In particular, Var-CNN achieves the highest $\pi_1$ and $\pi_5$ (50.3\% and 36.3\%), followed by NetCLR and DF. In contrast, multi-tab attacks such as BAPM, TMWF, and ARES yield significantly weaker performance, suggesting that the added complexity of modeling concurrent flows undermines robustness in open-world scenarios.

\subsection{Evaluation under Few-shot Setting}
\begin{table*}[]
\centering
\setlength{\belowcaptionskip}{0.1cm}
\caption{Performance of WF methods under Few-shot Setting}
\label{tab:few_shot}
\begin{tabular}{@{}l|cccc|cccc@{}}
\toprule
\multirow{2}{*}{\textbf{Method}} & \multicolumn{4}{|c|}{\textbf{Day 270}} & \multicolumn{4}{c}{\textbf{Day 480}} \\ \cline{2-5} \cline{6-9} 
 & \textbf{Acc.} & \textbf{Prc.} & \textbf{Rec.} & \textbf{F1} & \textbf{Acc.} & \textbf{Prc.} & \textbf{Rec.} & \textbf{F1} \\ \midrule
AWF & 0.3722 & 0.3723 & 0.3717 & 0.3625 & 0.2944 & 0.2867 & 0.2943 & 0.2829 \\
DF & 0.5897 & 0.6002 & 0.5897 & 0.5795 & 0.432 & 0.4738 & 0.4321 & 0.4264 \\
VarCNN & 0.7042 & 0.7090 & 0.7025 & 0.699 & 0.5412 & 0.5947 & 0.5411 & 0.5428 \\
TikTok & 0.6647 & 0.6715 & 0.6654 & 0.6586 & 0.5183 & 0.5457 & 0.5181 & 0.5134 \\
TF & 0.4914 & 0.4982 & 0.4914 & 0.4791 & 0.429 & 0.4279 & 0.4289 & 0.4162 \\
BAPM & 0.4916 & 0.5202 & 0.4927 & 0.483 & 0.3914 & 0.4027 & 0.3912 & 0.3680 \\
ARES & 0.7084 & 0.7422 & 0.7090 & 0.7097 & 0.6130 & 0.6541 & 0.6131 & 0.6103 \\
NetCLR & 0.2214 & 0.2419 & 0.2226 & 0.1945 & 0.2317 & 0.2377 & 0.2316 & 0.1935 \\
TMWF & 0.2594 & 0.4556 & 0.2616 & 0.2713 & 0.4419 & 0.4727 & 0.4419 & 0.4367 \\
RF & \textbf{0.8134} & \textbf{0.8211} & \textbf{0.814} & \textbf{0.8087} & \textbf{0.7004} & \textbf{0.7355} & \textbf{0.7003} & \textbf{0.7007} \\
Holmes & 0.6195 & 0.6302 & 0.6175 & 0.5989 & 0.4466 & 0.4771 & 0.4465 & 0.4268 \\
\bottomrule
\end{tabular}
\end{table*}

To evaluate robustness and few-shot learning under temporal drift, we construct a longitudinal dataset based on the March 2024 Tranco list, selecting 102 consistently accessible websites over a 16-month period (March 2024–July 2025). Traffic collection began on March 13, 2024 (Day 0) and continued at six subsequent checkpoints: Days 14, 30, 90, 150, 270, and 480. In total, 145,896 traces were gathered via Tor 0.4.8 from Singapore cloud servers. Models were initially trained on Day 0 data for 30 epochs and later fine-tuned with only 10 labeled samples per class from later checkpoints (Day 270 or Day 480).

As presented in Table~\ref{tab:few_shot}, RF and ARES achieve the strongest performance. RF maintains relatively high accuracy, achieving 0.8134 on Day 270 and 0.7004 on Day 480, while ARES shows a similar trend but experiences a sharper decline, reaching 0.6130 on Day 480. In contrast, AWF, BAPM, and NetCLR exhibit poor performance under few-shot conditions. Although TF is specifically designed for few-shot learning, its accuracy drops substantially, falling to 0.4290 on Day 480, and remains inferior to RF and ARES.

These results underscore the importance of robust feature extraction in scenarios with limited training data. Notably, general-purpose methods such as RF and ARES demonstrate greater resilience to temporal drift and outperform specialized few-shot approaches, highlighting their potential for long-term deployment.
\section{Discussion}

\noindent \textbf{The complexity of real-world scenarios precludes the universality of any single optimization strategy.}  
Our results show that no WF attack maintains strong performance across all conditions. Deep learning models achieve near-perfect accuracy in closed, defense-free settings but degrade under multi-label, open-world, or low-data constraints. ARES performs well with multi-label and mixed traffic but is limited by training scale and temporal coverage. Holmes is effective for early recognition but fails with traffic mixtures. Few-shot augmentation alleviates data scarcity but does not address defenses or multi-label challenges. Each method offers strengths, yet none is universally robust.  
This highlights the multidimensional nature of real-world settings: gains in one dimension often incur losses in others. CNNs fail under traffic drift due to reliance on static patterns; Transformer-based models demand extensive training data and become unstable in few-shot cases; augmented samples lack robustness against defenses. The central limitation of WF attacks is insufficient cross-scenario robustness—a challenge also observed in broader secure machine learning, where models collapse once underlying assumptions change.

\noindent \textbf{A comprehensive evaluation framework exposes limitations overlooked by traditional studies.}  
Our multi-scenario framework acts as a ``truth mirror'' for WF attacks. Prior work often claimed improvements based on narrow metrics, but broader evaluation reveals deeper flaws. DF models, for example, show steep error growth under traffic drift, exposing weak temporal robustness hidden by fixed train–test splits. ARES performs better than expected in open-world tests, suggesting multi-label modeling can reduce false positives—an insight absent in its original study. Few-shot augmentation improves closed-world accuracy but provides little benefit in open-world scenarios, as synthetic samples fail to capture unseen diversity.  
These findings illustrate how a comprehensive evaluation corrects overly optimistic conclusions. A model reporting 99\% accuracy in one scenario may drop to 50\% under slightly more realistic conditions—clearly inadequate for real attacks. This underscores the importance of unified benchmarks: only through systematic evaluation can WF attacks be compared fairly and their real applicability assessed.

\noindent \textbf{Future directions.}  
Future research should pursue three paths. First, multi-scenario joint optimization through multi-task or meta-learning may allow models to adapt across diverse conditions. Second, dynamic adversarial strategies inspired by game theory could enable attacks and defenses to adapt to each other in real time. Third, standardized datasets and protocols are essential: expanding the proposed framework with more public datasets and evaluation dimensions would foster a unified benchmarking platform. Finally, integrating WF attacks with complementary techniques such as traffic manifold modeling or anomaly detection may improve resilience in unseen scenarios.

\section{Conclusion}

In this paper, we present a systematic and multidimensional evaluation of existing WF attacks across six realistic challenges: defense mechanisms, traffic drift, multi-tab browsing, early-stage detection, open-world scenarios, and few-shot settings. The experimental results demonstrate that although certain methods achieve high accuracy in isolated scenarios, their performance degrades significantly under complex conditions, underscoring the lack of cross-scenario robustness. These findings not only reveal the limitations of prior research constrained by single-perspective evaluations but also highlight the resilience of anonymity networks in the presence of diverse protections and dynamic environments. Future work should further explore joint optimization through multi-task or meta-learning, dynamic adversarial strategies, and standardized benchmark datasets, thereby advancing WF attack research toward greater practicality and robustness.

\renewcommand\refname{\zihao{5}\textbf{References}}
\bibliographystyle{IEEEtran} 
\bibliography{references}

\begin{thebibliography}{10}
\providecommand{\url}[1]{#1}
\csname url@samestyle\endcsname
\providecommand{\newblock}{\relax}
\providecommand{\bibinfo}[2]{#2}
\providecommand{\BIBentrySTDinterwordspacing}{\spaceskip=0pt\relax}
\providecommand{\BIBentryALTinterwordstretchfactor}{4}
\providecommand{\BIBentryALTinterwordspacing}{\spaceskip=\fontdimen2\font plus
\BIBentryALTinterwordstretchfactor\fontdimen3\font minus \fontdimen4\font\relax}
\providecommand{\BIBforeignlanguage}[2]{{%
\expandafter\ifx\csname l@#1\endcsname\relax
\typeout{** WARNING: IEEEtran.bst: No hyphenation pattern has been}%
\typeout{** loaded for the language `#1'. Using the pattern for}%
\typeout{** the default language instead.}%
\else
\language=\csname l@#1\endcsname
\fi
#2}}
\providecommand{\BIBdecl}{\relax}
\BIBdecl

\bibitem{tor2024users}
\BIBentryALTinterwordspacing
``Users - tor metrics,'' 2024. [Online]. Available: \url{https://metrics.torproject.org/userstats-relay-country.html}
\BIBentrySTDinterwordspacing

\bibitem{dingledine2004tor}
R.~Dingledine, N.~Mathewson, and P.~Syverson, ``Tor: The second-generation onion router,'' Naval Research Lab Washington DC, Tech. Rep., 2004.

\bibitem{wang2014effective}
T.~Wang, X.~Cai, R.~Nithyanand, R.~Johnson, and I.~Goldberg, ``Effective attacks and provable defenses for website fingerprinting,'' in \emph{23rd USENIX Security Symposium}, 2014, pp. 143--157.

\bibitem{hayes2016k}
J.~Hayes and G.~Danezis, ``k-fingerprinting: A robust scalable website fingerprinting technique,'' in \emph{25th USENIX Security Symposium}, 2016, pp. 1187--1203.

\bibitem{wang2018deep}
K.~Wang, M.~Yang, W.~Yang, and Y.~Yin, ``Deep correlation structure preserved label space embedding for multi-label classification,'' in \emph{Asian Conference on Machine Learning}.\hskip 1em plus 0.5em minus 0.4em\relax PMLR, 2018, pp. 1--16.

\bibitem{xu2018multi}
Y.~Xu, T.~Wang, Q.~Li, Q.~Gong, Y.~Chen, and Y.~Jiang, ``A multi-tab website fingerprinting attack,'' in \emph{Proceedings of the 34th Annual Computer Security Applications Conference}, 2018, pp. 327--341.

\bibitem{rimmer2018automated}
V.~Rimmer, D.~Preuveneers, M.~Juarez, T.~Van~Goethem, and W.~Joosen, ``Automated website fingerprinting through deep learning,'' in \emph{NDSS}, 2018.

\bibitem{juarez2014critical}
M.~Juarez, S.~Afroz, G.~Acar, C.~Diaz, and R.~Greenstadt, ``A critical evaluation of website fingerprinting attacks,'' in \emph{Proceedings of the 2014 ACM SIGSAC Conference on Computer and Communications Security}, 2014, pp. 263--274.

\bibitem{wang2020high}
T.~Wang, ``High precision open-world website fingerprinting,'' in \emph{2020 IEEE Symposium on Security and Privacy (SP)}.\hskip 1em plus 0.5em minus 0.4em\relax IEEE, 2020, pp. 152--167.

\bibitem{torversion}
\BIBentryALTinterwordspacing
``Support tech for freedom and human rights.'' 2019. [Online]. Available: \url{https://blog.torproject.org/giving-tuesday-support-tech-freedom-and-human-rights/}
\BIBentrySTDinterwordspacing

\bibitem{deng2023robust}
X.~Deng, Q.~Yin, Z.~Liu, X.~Zhao, Q.~Li, M.~Xu, K.~Xu, and J.~Wu, ``Robust multi-tab website fingerprinting attacks in the wild,'' in \emph{2023 IEEE Symposium on Security and Privacy (SP)}.\hskip 1em plus 0.5em minus 0.4em\relax IEEE Computer Society, 2023, pp. 1005--1022.

\bibitem{jin2023transformer}
Z.~Jin, T.~Lu, S.~Luo, and J.~Shang, ``Transformer-based model for multi-tab website fingerprinting attack,'' in \emph{Proceedings of the 2023 ACM SIGSAC Conference on Computer and Communications Security}, 2023, pp. 1050--1064.

\bibitem{guan2021bapm}
Z.~Guan, G.~Xiong, G.~Gou, Z.~Li, M.~Cui, and C.~Liu, ``Bapm: Block attention profiling model for multi-tab website fingerprinting attacks on tor,'' in \emph{Annual Computer Security Applications Conference}, 2021, pp. 248--259.

\bibitem{sirinam2019triplet}
P.~Sirinam, N.~Mathews, M.~S. Rahman, and M.~Wright, ``Triplet fingerprinting: More practical and portable website fingerprinting with n-shot learning,'' in \emph{Proceedings of the 2019 ACM SIGSAC Conference on Computer and Communications Security}, 2019, pp. 1131--1148.

\bibitem{jansen2024measurement}
R.~Jansen, R.~Wails, and A.~Johnson, ``A measurement of genuine tor traces for realistic website fingerprinting,'' \emph{arXiv preprint arXiv:2404.07892}, 2024.

\bibitem{cherubin2022online}
G.~Cherubin, R.~Jansen, and C.~Troncoso, ``Online website fingerprinting: Evaluating website fingerprinting attacks on tor in the real world,'' in \emph{31st USENIX Security Symposium (USENIX Security 22)}, 2022, pp. 753--770.

\bibitem{liberatore2006inferring}
M.~Liberatore and B.~N. Levine, ``Inferring the source of encrypted http connections,'' in \emph{Proceedings of the 13th ACM conference on Computer and communications security}, 2006, pp. 255--263.

\bibitem{panchenko2011website}
A.~Panchenko, L.~Niessen, A.~Zinnen, and T.~Engel, ``Website fingerprinting in onion routing based anonymization networks,'' in \emph{Proceedings of the 10th annual ACM workshop on Privacy in the electronic society}, 2011, pp. 103--114.

\bibitem{cai2012editdistance}
\BIBentryALTinterwordspacing
X.~Cai, X.~C. Zhang, B.~Joshi, and R.~Johnson, ``Touching from a distance: website fingerprinting attacks and defenses,'' in \emph{the {ACM} Conference on Computer and Communications Security, CCS'12, Raleigh, NC, USA, October 16-18, 2012}, T.~Yu, G.~Danezis, and V.~D. Gligor, Eds.\hskip 1em plus 0.5em minus 0.4em\relax {ACM}, 2012, pp. 605--616. [Online]. Available: \url{https://doi.org/10.1145/2382196.2382260}
\BIBentrySTDinterwordspacing

\bibitem{sirinam2018deep}
P.~Sirinam, M.~Imani, M.~Juarez, and M.~Wright, ``Deep fingerprinting: Undermining website fingerprinting defenses with deep learning,'' in \emph{Proceedings of the 2018 ACM SIGSAC Conference on Computer and Communications Security}, 2018, pp. 1928--1943.

\bibitem{bhat2019varcnn}
S.~Bhat, D.~Lu, A.~Kwon, and S.~Devadas, ``Var-cnn: A data-efficient website fingerprinting attack based on deep learning,'' \emph{Proceedings on Privacy Enhancing Technologies}, vol.~4, pp. 292--310, 2019.

\bibitem{rahman2019tik}
M.~S. Rahman, P.~Sirinam, N.~Mathews, K.~G. Gangadhara, and M.~Wright, ``Tik-tok: The utility of packet timing in website fingerprinting attacks,'' \emph{Proceedings on Privacy Enhancing Technologies}, vol.~3, pp. 5--24, 2020.

\bibitem{bahramali2023realistic}
A.~Bahramali, A.~Bozorgi, and A.~Houmansadr, ``Realistic website fingerprinting by augmenting network traces,'' in \emph{Proceedings of the 2023 ACM SIGSAC Conference on Computer and Communications Security}, 2023, pp. 1035--1049.

\bibitem{shen2023rf}
M.~Shen, K.~Ji, Z.~Gao, Q.~Li, L.~Zhu, and K.~Xu, ``Subverting website fingerprinting defenses with robust traffic representation,'' in \emph{32nd USENIX Security Symposium (USENIX Security 23)}, 2023, pp. 607--624.

\bibitem{deng2024holmes}
X.~Deng, Q.~Li, and K.~Xu, ``Robust and reliable early-stage website fingerprinting attacks via spatial-temporal distribution analysis,'' in \emph{Proceedings of the 2024 on ACM SIGSAC Conference on Computer and Communications Security}, 2024, pp. 1997--2011.

\bibitem{schroff2015facenet}
F.~Schroff, D.~Kalenichenko, and J.~Philbin, ``Facenet: A unified embedding for face recognition and clustering,'' in \emph{Proceedings of the IEEE conference on computer vision and pattern recognition}, 2015, pp. 815--823.

\bibitem{vaswani2017attention}
A.~Vaswani, N.~Shazeer, N.~Parmar, J.~Uszkoreit, L.~Jones, A.~N. Gomez, {\L}.~Kaiser, and I.~Polosukhin, ``Attention is all you need,'' in \emph{Advances in neural information processing systems}, 2017, pp. 5998--6008.

\bibitem{lin2013network}
M.~Lin, Q.~Chen, and S.~Yan, ``Network in network,'' \emph{arXiv preprint arXiv:1312.4400}, 2013.

\bibitem{khosla2020supervised}
P.~Khosla, P.~Teterwak, C.~Wang, A.~Sarna, Y.~Tian, P.~Isola, A.~Maschinot, C.~Liu, and D.~Krishnan, ``Supervised contrastive learning,'' \emph{Advances in neural information processing systems}, vol.~33, pp. 18\,661--18\,673, 2020.

\bibitem{wang2017walkie}
T.~Wang and I.~Goldberg, ``Walkie-talkie: An efficient defense against passive website fingerprinting attacks,'' in \emph{26th USENIX Security Symposium}, 2017, pp. 1375--1390.

\bibitem{yang2021cade}
L.~Yang, W.~Guo, Q.~Hao, A.~Ciptadi, A.~Ahmadzadeh, X.~Xing, and G.~Wang, ``Cade: Detecting and explaining concept drift samples for security applications,'' in \emph{30th USENIX Security Symposium (USENIX Security 21)}, 2021, pp. 2327--2344.

\bibitem{oh2021gandalf}
S.~E. Oh, N.~Mathews, M.~S. Rahman, M.~Wright, and N.~Hopper, ``Gandalf: Gan for data-limited fingerprinting,'' \emph{Proceedings on Privacy Enhancing Technologies}, vol. 2021, no.~2, 2021.

\end{thebibliography}

\end{document}